# Controlling the Flow of Information in Optical Metrology

Maximilian Weimar[1*], Huanli Zhou[2*], Luca Neubacher[1*], Thomas A. Grant[2*], Jakob Hüpfl[1], Kevin F. MacDonald[2], Stefan Rotter[1†], and Nikolay I. Zheludev[2,3‡]

[1] *Institute for Theoretical Physics, Vienna University of Technology (TU Wien), A-1040 Vienna, Austria*

[2] *Optoelectronics Research Centre, University of Southampton, Highfield, Southampton, SO17 1BJ, UK*

[3] *Hagler Institute for Advanced Study, Texas A&M University, College Station, Texas, 77843, USA*

**ABSTRACT:**

Optical metrology has progressed beyond the Abbe–Rayleigh limit, unlocking (sub)atomic precision by leveraging nonlinear phenomena, statistical accumulation, and AI estimators trained on measurand variations. Here, we show that Fisher information, which defines the fundamental precision limit, can be viewed as a physical entity that propagates through space, and we derive a wave equation for sensitivity fields describing its flow, which can resonate, diffract, and interfere. We reveal how material composition, geometry, and environmental design dictate where information is generated and how it travels, analogous to antennas and metasurfaces sculpting electromagnetic energy. Plasmonic and dielectric resonances enhance information flow, while gratings and near-field structures reshape radiation patterns. This perspective reframes metrology as a discipline in which resolution can be engineered by tailoring information sources and flow for applications in atomic-scale diagnostics and beyond, including optimisation of Light Detection and Ranging (LiDAR), remote sensing, and radar technologies.

---

*Contributed equally to this work

†E-mail: stefan.rotter@tuwien.ac.at

‡E-mail: zheludev@soton.ac.uk

## Introduction: optical metrology and information

Optical metrology has a broad scope of applications in smart manufacturing, precise instrumentation, nanotechnology, bioimaging and materials research. While sophisticated interferometry is used for positional metrology of macroscopic objects, such as in LIGO gravitational wave detectors[1] and laser encoders for motion control stages, dimensional and positional optical metrology of nanoscale objects requires different approaches. To break the diffraction limit of optical resolution, nonlinear fluorescence microscopy techniques exploit the depletion of emission from biological samples tagged with fluorophores[2,3], and statistical single-molecule localization microscopies detect the positions of sparse emitters by averaging information from thousands of images[4,5]. Both can achieve nanometric resolution but suffer from long acquisition times and require complex and expensive instrumentation.

A variety of approaches based on interferometric light and evanescent field scattering and beam deflection have been reported for tracking the position of isolated (often optically trapped) nanoparticles with sub-nanometric precision[6-11]. A range of techniques have also emerged for translational position measurement and localisation, again with nano- to sub-nanometric precision, based on the use of light fields structured at sub-wavelength scales (by metasurfaces, plasmonic nanostructures, and spatial light modulators) to achieve high sensitivity of scattered light profiles to object positions[12-15].

It has also been shown recently that high-precision dimensional and localization measurements can be achieved by analysing scattered intensity patterns in the simple setting of a laboratory microscope, through the use of a neural network as a measurement estimator (trained – i.e. provided with prior information – either on a number of similar objects or indeed the target object itself with controlled, in-situ variation of the desired parameter). This approach has been applied to single-shot multi-parameter dimensional measurements (length, width, separation) on subwavelength slits and nanorod dimers[16,17], and to nanowire localization (displacement) measurements with precision reaching sub-atomic length scales beyond 100 picometers[18,19], up to a million measurement per second[1] (enabling 'real-time' tracking of the object's thermal motion[20]). In this context, the use of a topologically structured incident light field containing singularities improves measurement precision manifold by enhancing the dependence of scattered intensity profiles on object position, i.e. their information content[21].

We argue that further improvements in all kinds of optical scattering metrology can be achieved by better understanding of the information aspect. Here, a central concept is Fisher information[22] (FI), which determines the Cramér-Rao bound – the ultimate limit on the precision with which a parameter, such as the position or size of a scatterer, can be estimated from a given measurement. Recent advances have demonstrated that shaping the input wavefront[21,23] or optimizing the detection strategy[24-26] to maximize the FI leads to substantial improvements in metrological precision.

In a recent conceptual advance, it was shown that the FI does not merely quantify a measurement's resolution but can be understood as a quantity that is carried by electromagnetic waves as they propagate through space[27]. In this view, the FI emerges when a light field interacts with an object of interest and is carried by the scattered field through the surrounding medium toward the detector. This interpretation, grounded in a continuity equation akin to those found in classical field theories, reveals that FI is conserved in dissipation-free regions, much like energy. Here, we will demonstrate a fundamental consequence of this formulation, which is that the flow of FI inherits the full wave-like character of light itself: it can resonate, diffract, and, perhaps most surprisingly, interfere.

As we will demonstrate explicitly in this work, this phenomenon of "information interference" has profound implications for optical design. Specifically, we will show that the FI can be resonantly enhanced in structured environments, and that its flow can be steered, analogously to beam steering, through appropriate design of a target's environment. Moreover, we will show that the FI flow depends not only on the shape and size of scattering objects, but also on their material composition. Our findings suggest a new paradigm for optical design: one in which information is treated not merely as a quantity to be passively extracted, but as a wave-like entity to be actively guided and controlled through interference and choice of materials.

**Fisher information and light scattering**

Our starting point will be to demonstrate that the creation of Fisher information in a scattering process crucially depends on the resonant nature of the scattering target. We consider light scattering from a small object in free space[28] (see Fig. 1) described by the coefficients $c_{in}$ and $c_{scat}$, representing the incident and scattered electric field in a suitable basis such as vector spherical harmonics. As the field scatters from the object and propagates to a detector located in the far-field, it carries a certain amount of Fisher information $J$ about any parameter $\theta$ that characterises the object, such as its position or its angular orientation. In general, when this parameter $\theta$ is estimated from measurement data, the precision of this estimation is constrained by the Cramér-Rao bound $Var(\hat{\theta}) \geq \frac{1}{J}$, where $\hat{\theta}$ is the estimator of the parameter $\theta$. Here, the Fisher information $J = \int p(x;\theta)[\partial_\theta \, log\,(p(x;\theta))]^2\, dx$, depends on the probability density function $p(x;\theta)$ describing the measurement data and its changes with respect to the parameter of interest $\theta$. When data is collected by probing a particle with coherent light at frequency $\omega$, the maximum amount of Fisher information one can extract through an optimal measurement procedure[12] is equal to the quantum Fisher information and is given by $J = 4(\partial_\theta c_{scat})^\dagger \partial_\theta c_{scat}/\hbar\omega$. To unveil the resonant behavior in this expression for Fisher information in the far-field, it can be connected to a particle's scattered power $P_{scat} \sim c_{scat}^\dagger c_{scat}$, in which resonances originating from Mie or plasmonic scattering emerge as pronounced peaks. Following some simple algebraic manipulations, one may verify that $J = \left[2\partial_\theta^2 P_{scat} - 4Re((\partial_\theta^2 c_{scat}^\dagger)c_{scat})\right]/\hbar\omega$. The first term on the right-hand side features the total scattered power $P_{scat}$ and is only dominant when neither the phase nor the power distribution of scattered light carries Fisher information (see Methods). This occurs, for example, when a subwavelength particle is located at a node of a standing wave or at a radial node of a Bessel beam where $c_{scat} = 0$. For plane-wave illumination and when the parameter $\theta$ of interest is the particle's position in free space, the first term on the right-hand side vanishes and the Fisher information $J \approx -4Re[(\partial_{r_i}^2 c_{scat}^\dagger)c_{scat}]/\hbar\omega$. This expression suggests that strong scattering should give rise to large Fisher information, but only if the vectors $c_{scat}$ and $\partial_{r_i}^2 c_{scat}$ are well aligned, which is found to be the case for small particles in which the excitation of a dipole component dominates. In this case, the Fisher information is directly proportional to the scattered power $J \propto kP_{scat}$ (see Methods and Fig. 2e). However, the proportionality with wavenumber $k$ does not hold universally. As shown in the Methods section and in Fig. 2f, the presence of more than one multipolar contribution to the scattered field[29,30] breaks the scaling: Fisher information then depends nonlinearly on $k$ as the second-derivative operator $\partial_{r_i}^2$ acts differently on different multipole orders. Therefore, although scattered light is the carrier of information on target position, the total scattered power and Fisher information may have different spectral dispersions.

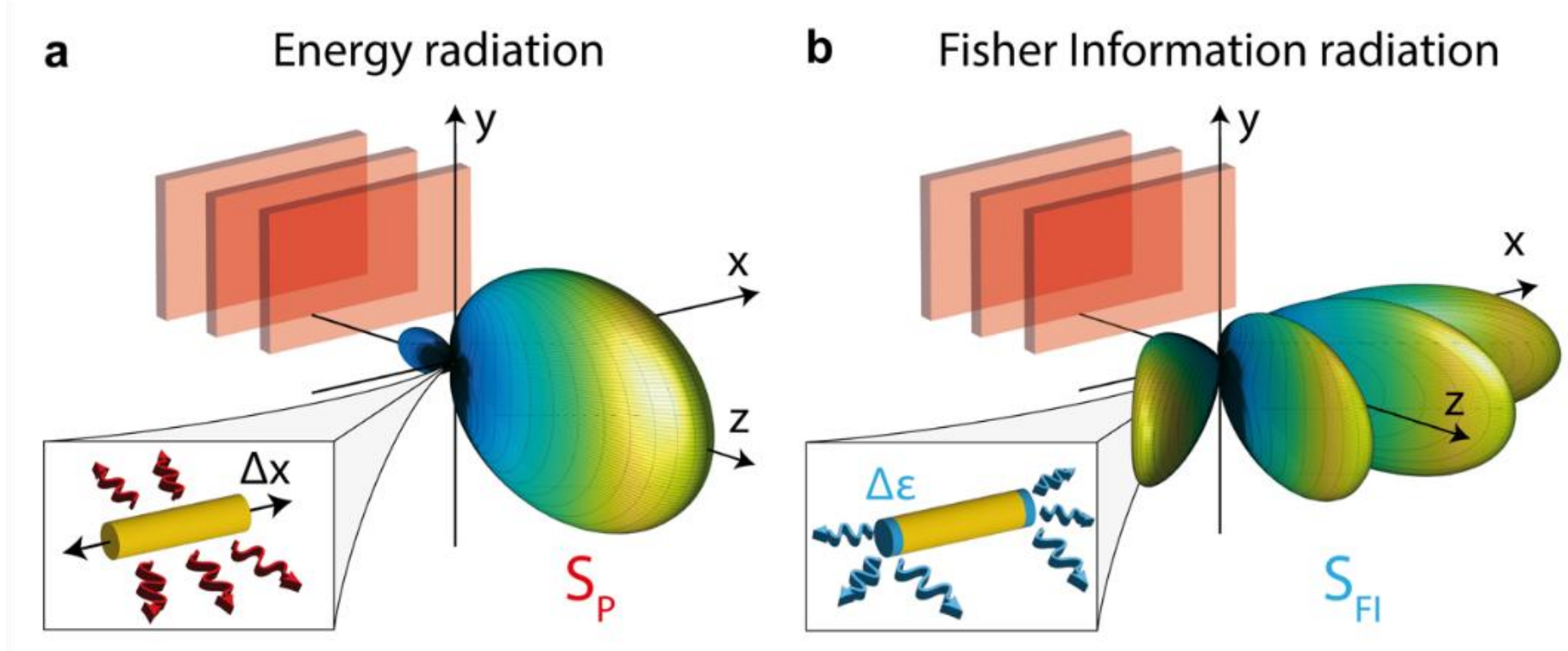

**Fig. 1. Optical metrology with scattered light.** (a, b) An incoming plane wave impinges on a measurement target (e.g. a nanorod), generating a scattered wave. (a) Energy radiated into the far-field is described by the Poynting vector $\boldsymbol{S}_P$. (b) The radiation pattern of the Fisher information flux vector $\boldsymbol{S}_{FI}$ is generally different from $\boldsymbol{S}_P$, as the sources of Fisher information and scattered energy are different in nature. For measurements of the $x$-position of the nanorod illustrated here, the information radiation pattern is defined by interference between Fisher information sources with opposing phase located at the opposing ends of the target (the blue shaded regions in the inset to (b)).

Interestingly, these calculations also show that the proportionality factor in $J \propto kP_{scat}$ for a single multipolar resonance depends strongly on the probe beam's polarization with respect to the parameter of interest (see Methods). For the case considered in Figs. 2a, c, e, where most of the scattering occurs due to the electric dipole, an incoming plane wave propagating along $\hat{\boldsymbol{z}}$ and polarized along $\hat{\boldsymbol{x}}$ produces twice as much Fisher information on the particle's $y$-displacement and seven times as much on its $z$-displacement, than on its $x$-displacement. This strong dependence suggests that simply choosing the right incoming polarization with respect to the parameter of interest can provide remarkable advantage in a measurement process.

## Energy and information flow in optical metrology

The time averaged Fisher information arriving at a detector with area $A$ and surface-normal $\hat{\boldsymbol{n}}$ can be expressed in terms of Maxwellian electrodynamics as $J(\theta) = \frac{4}{T\hbar\omega}\int_{t-\frac{T}{2}}^{t+\frac{T}{2}}\int_A \left(\partial_\theta \boldsymbol{E}(\tilde{t}) \times \partial_\theta \boldsymbol{H}(\tilde{t})\right) d\hat{\boldsymbol{n}} d\tilde{t}$, where $T$ is the light wave period. From here, the Fisher information flow of light can be defined as[27] $\boldsymbol{S}_{FI} = \frac{2}{\hbar\omega} Re(\partial_\theta \boldsymbol{E}_\omega^* \times \partial_\theta \boldsymbol{H}_\omega)$, and instructively compared with the Poynting vector for the flow of light energy $\boldsymbol{S}_P = \frac{1}{2} Re(\boldsymbol{E}_\omega^* \times \boldsymbol{H}_\omega)$. Figure 2 compares energy flows for plasmonic and dielectric nanospheres under plane wave illumination, and corresponding FI flows for displacement measurements in the transverse and longitudinal directions. It is clear to see that energy and Fisher information do not necessarily propagate in the same direction. Indeed, the Fisher information flow may even be zero in the dominant direction of light propagation, as indicated by the red dashed lines of reflection symmetry in Figs. 2c, d. Based on these information flow patterns, the above-mentioned ratio of 1:2:7 in Fisher information for *x*-, *y*- and *z*-displacements of the gold nanosphere can be intuitively understood: Information is lowest when the incoming plane wave polarized along $\hat{\boldsymbol{x}}$ excites a dipole radiation pattern whose main lobes coincide with the *y*-*z* symmetry plane of the problem, into which no Fisher information can flow. For $y$-displacement, where this is not the case, there is twice as much Fisher information. Similar arguments apply to $z$-displacement, but motion in this direction also shifts the phase of the induced dipole. This phase shift applies uniformly to the entire scattered field and produces an additional term for the Fisher information, leading to the largest scaling factor (×7, relative to the $x$-displacement case). For the silicon particle (Figs. 2b, d, f), the scattering behaviour is more complicated as several resonances arise. Still, under the constraint that an overlap between these resonances needs to be considered, the same arguments can be applied: The behaviour of the magnetic dipole is analogous to that of the electric dipole, except that the effective polarisation is rotated. For the quadrupolar resonances, the Fisher information for $x$ and $y$ displacements becomes equal, because the scattered field vanishes in the corresponding symmetry planes for both parameters.

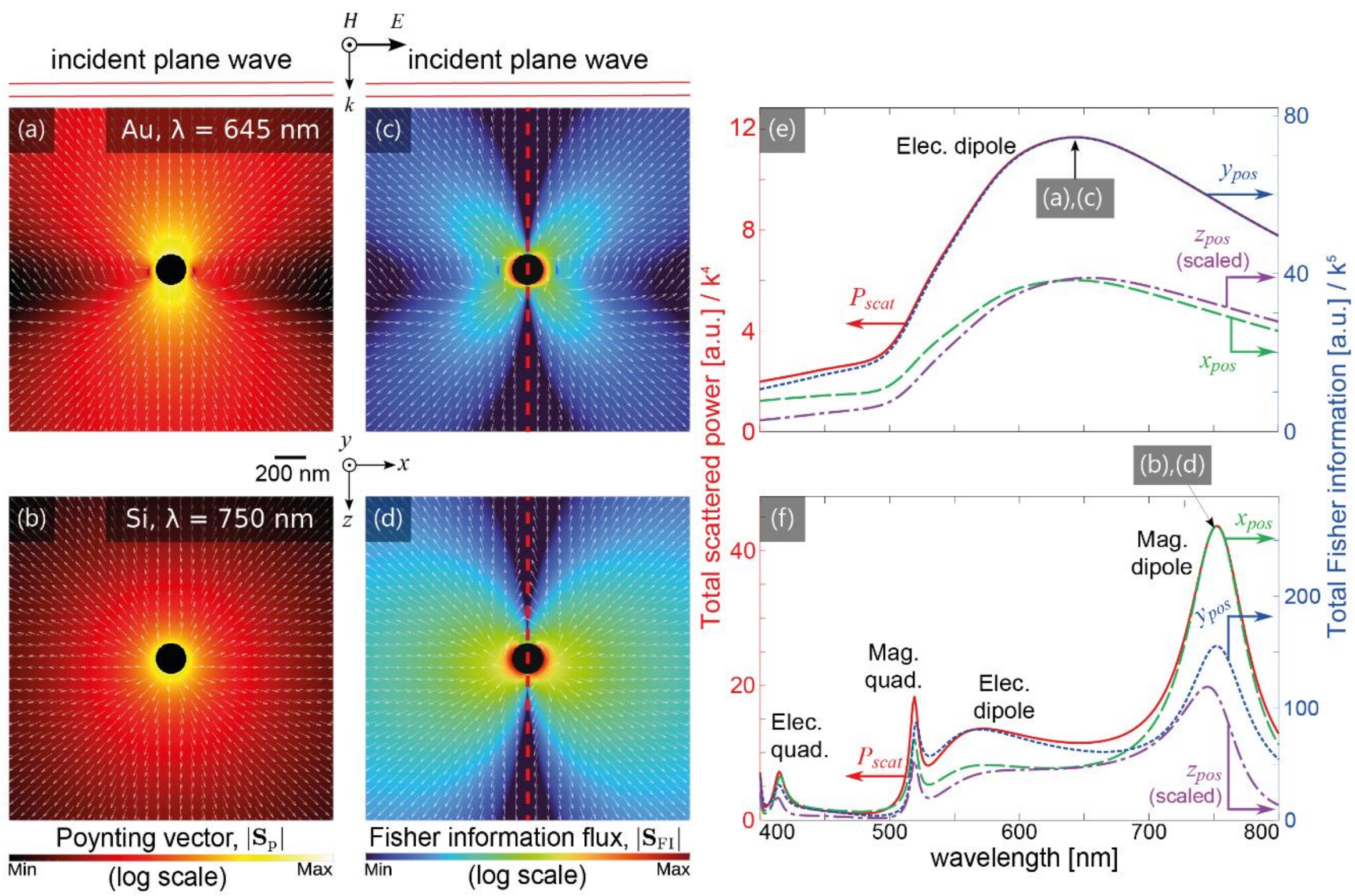

**Fig. 2. Energy and Fisher information flows.** (a,b) Cross-sectional energy flow [Poynting vector] maps of the scattered field around nanospheres of 95 nm radius made from (a) gold and (b) silicon, under plane wave illumination along the $z$-direction, with polarization in the $x$-direction. (c,d) Fisher information flow maps for $x$-displacements of the nanospheres shown in (a,b), respectively. In all cases, the amplitudes of flow vectors are color-coded, and their directions are represented by the overlaid white arrows. The red dashed lines in (c,d) denote symmetry planes where no Fisher information can flow. (e,f) Spectral densities of total scattered power ($P_{scat}$) and total Fisher information (FI) for $x$-, $y$- and $z$-displacements of the spheres shown in (a,c) and (b,d), respectively. For ease of comparison, the FI for $z$-displacements is scaled down by a factor of 7. Labels and black arrows denote the FI-peak wavelengths at which the field maps in (a,c) and (b,d) are plotted: specifically, 645 nm for the gold nanosphere; 750 nm for silicon. Peaks are also labelled with the dominant multipolar resonance [quad. = quadrupole].

## Wave-like nature of Fisher information: interference and diffraction

The difference between the flows of scattered light energy and Fisher information (Figs 2a vs. 2c, and 2b vs. 2d) may also be understood from the difference in their sources. According to Huygens, two points in space reached by an incident wavefront at the same time emit secondary waves that are in phase. In Young's double slit experiment, *constructive interference* of waves on the symmetry line separating the slits is an example of this. Does Huygens' principle also hold for the flow of FI, and what quantity is interfering in that case? We argue here that the wave-like behaviour of FI propagation resides in the 'sensitivity fields' $\partial_\theta \boldsymbol{E}$ and $\partial_\theta \boldsymbol{H}$, which describe how the physical fields $\boldsymbol{E}$ and $\boldsymbol{H}$ change when the measured parameter is varied[31].

In the same way as the electric and magnetic currents $\boldsymbol{j}^e$, $\boldsymbol{j}^m$ in an antenna emit the electric and magnetic fields, effective electric and magnetic currents can be defined[27]: $\boldsymbol{j}_{eff}^{e} = \partial_t \boldsymbol{E} \partial_\theta \epsilon + \partial_\theta \boldsymbol{j}$ and $\boldsymbol{j}_{eff}^{m} = \partial_t H \partial_\theta \mu$, which emit the sensitivity fields $\partial_\theta \boldsymbol{E}$ and $\partial_\theta \boldsymbol{H}$ . Here $\epsilon$, μ, and $\boldsymbol{j}$ correspond to the space-dependent permittivity, permeability and free current, respectively. These effective electric and magnetic currents only exist in regions where at least one of these quantities is dependent on $\theta$, i.e. $\partial_\theta \epsilon, \partial_\theta \mu, \partial_\theta \boldsymbol{j} \neq 0$. We will henceforth only consider systems with constant permeability $\mu_0$ with which, starting from Maxwell's equations, it can be shown that the electric sensitivity fields $\partial_\theta \boldsymbol{E}$ are solutions of the wave equation for the sensitivity fields that control the flow of Fisher information:

$$\nabla \times \nabla \times [\partial_\theta \boldsymbol{E}] + \mu_0 \epsilon \partial_t^2 [\partial_\theta \boldsymbol{E}] = \boldsymbol{Q}.$$

Here, the right-hand side $\boldsymbol{Q} = -\mu_0 \partial_t \boldsymbol{j}_{eff}^{e}$, is non-zero only in those regions where the material is affected by a variation of the parameter $\theta$. (A similar expression can also be derived for the magnetic sensitivity field.) The left-hand side of the equation for the sensitivity field is the same as that of the wave equation for the fields themselves $\nabla \times \nabla \times \boldsymbol{E} + \mu\epsilon \partial_t^2 \boldsymbol{E} = 0$, the only difference being the source term $\boldsymbol{Q}$ on the right. Correspondingly, the sensitivity fields can be found as a superposition of contributions originating at the sources $\boldsymbol{Q}$ as: $\partial_\theta \boldsymbol{E} = \int d^3 r' dt' G(r, r', t, t') \boldsymbol{Q}(r', t')$. Here the (generally dyadic) Green's function is a fundamental solution to $\nabla \times \nabla \times G(r, r', t, t') + \mu\epsilon \partial_t^2 G(r, r', t, t') = \delta(r - r')\delta(t - t')$. Moreover, it follows from here that Fisher information propagates with the same speed of light as electromagnetic energy.

Applying these findings to a gold nanorod illuminated by a plane wave, where the parameter of interest is $\theta = x_{pos}$, as illustrated in Fig. 3, the effective electric currents are localised at the two ends of the nanorod, and no effective magnetic currents are present. This creates a double-source for Fisher information, but while the effective sources $\boldsymbol{j}_{eff}^{e} = \partial_t \boldsymbol{E} \partial_x \epsilon$ have equal magnitude, they are in antiphase since $\partial_x \epsilon$ changes sign between regions vacated and newly occupied by the nanorod's displacement. In contrast to the in-phase nature of light sources in Young's double slit experiment, this $\pi$ phase difference between the information sources leads to *destructive interference* of Fisher information flow along the symmetry line perpendicular to the nanorod (Figs. 3a,b). As such, a detector subtending a small angle about this symmetry axis will not readily, let alone optimally, detect changes of scattered intensity revealing/resolving small displacements in the lateral position of the nanowire. To overcome this deficiency, the FI flow pattern can be manipulated simply by changing the direction of the incident wavefront (Fig. 3c,d). At certain angles, obliquely incident illumination will introduce a phase delay in excitation between the two ends of the nanorod that offsets the $\pi$-phase difference between the effective information currents generated at those points, yielding constructive interference and thereby a FI flow maximum in the direction of the detector.

In addition to Huygens' principle for the sensitivity fields, a reciprocity relation can be established:

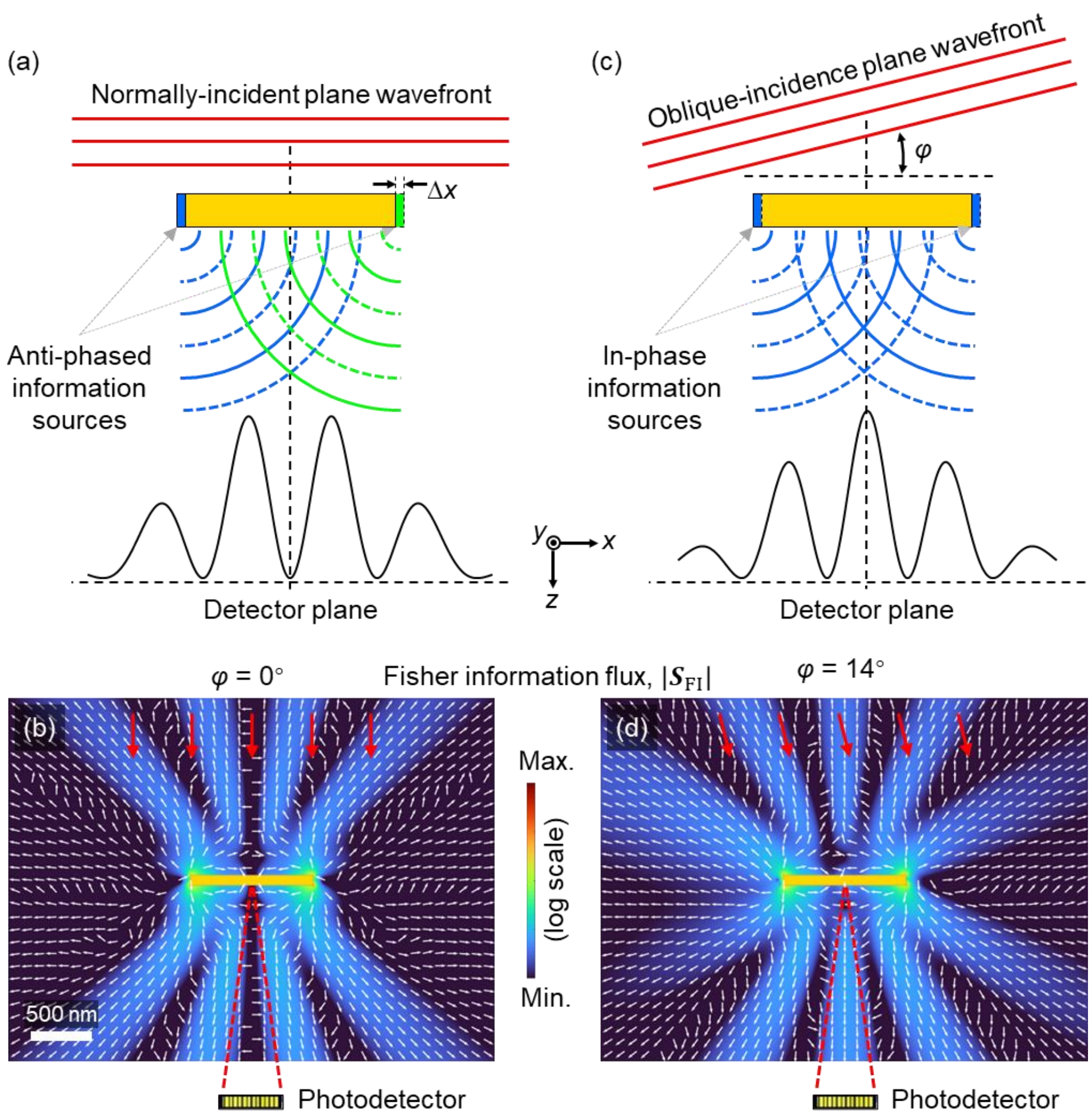

**Fig. 3. FI sources and FI interference in nanowire displacement metrology.** (a,b) Schematic illustrations of the fact that: (a) under normally-incident illumination, the FI sources at the edges of an object undergoing a lateral $\Delta x$ displacement [perpendicular to the propagation direction $z$ of plane wave illumination] have opposing phase, denoted by blue and green colors, yielding destructive interference of the sensitivity fields and an information minimum along the $x = 0$ direction; (b) Oblique incidence illumination at certain angles produces in-phase information sources, and thereby an information maximum along $x = 0$. (c,d) Corresponding cross-sectional FI flow maps pertaining to a $\Delta x = 2$ nm displacement of a 1 µm wide, 80 nm thick gold nanowire [of infinite extent in the $y$ direction] under plane wave illumination at a wavelength $\lambda = 500$ nm, at incident angles (c) $\varphi = 0°$, and (d) $\varphi = 14°$. On these two-dimensional maps, the amplitude of information flow vectors is color-coded and their directions are represented by the overlaid arrows.

From classical electrodynamics we know that when there are two localized currents $\boldsymbol{j}_1$ and $\boldsymbol{j}_2$, which produce electric fields $\boldsymbol{E}_1$ and $\boldsymbol{E}_2$ and magnetic fields $\boldsymbol{H}_1$ and $\boldsymbol{H}_2$, Lorentz reciprocity implies that $\int dV(\boldsymbol{j}_1 \cdot \boldsymbol{E}_2 - \boldsymbol{j}_2 \cdot \boldsymbol{E}_1) = 0$. As this relation is directly connected to the symmetry of the (time-independent) Green's function $G(\boldsymbol{r}, \boldsymbol{r}') = G^{\top}(\boldsymbol{r}', \boldsymbol{r})$, it can be shown that the corresponding relation for the sensitivity fields and effective currents is an analogue of Lorentz reciprocity in classical non-magnetic media, which holds when the fields and currents are replaced by sensitivity fields and effective currents $\int dV\big(\boldsymbol{j}^{e}_{eff,1}\cdot[\partial_\theta \boldsymbol{E}_2]-\boldsymbol{j}^{e}_{eff,2}\cdot[\partial_\theta \boldsymbol{E}_1]+\boldsymbol{j}^{mag}_{eff,1}\cdot[\partial_\theta \boldsymbol{H}_2]-\boldsymbol{j}^{mag}_{eff,2}\cdot[\partial_\theta \boldsymbol{H}_1]\big)=0$. In other words, the projection of the sensitivity fields generated at point 1 onto the effective currents at point 2 is equal to the projection of the sensitivity fields generated at point 2 onto the effective currents at point 1. In the same way as the reciprocity of many scattering systems is routinely used to reduce the complexity of physical measurements and numerical simulations, similar considerations can be applied to Fisher information. For instance, inverse design processes with Fisher information-based objectives, such as maximising or controlling information flow along certain spatial directions, will benefit from reciprocity: In the Methods section, we show how the widely used adjoint-variable method[32,33] extends naturally to objectives involving sensitivity fields. This extension relies on the symmetry of the underlying differential operator, which governs the propagation of both the physical fields and the associated sensitivity fields. As a result, this framework shall enable computationally efficient inverse design of structures that tailor and optimise Fisher information flow, such as directional radiating antennas.

Another powerful example of the quasi–wave nature of Fisher information is the 'Information Talbot effect', illustrated in Fig. 4. The classical optical Talbot effect is a diffraction phenomenon: When a plane wave is incident upon a diffraction grating of period $a$, the field pattern created by the grating is repeated at regular distances from the grating plane, equal to multiples of the Talbot length $z_T \approx \frac{2a^2}{\lambda}$. The repeated field patterns contain concentrations of energy density aligned with the grating slits (Fig. 4a), resembling the field structure in the nearfield of the grating (referred to as self-images of the grating). Taking the $x$-position of the grating as the parameter of interest, Fig. 4b shows maps of Fisher Information density $u_{FI}=(\epsilon|\partial_\theta \boldsymbol{E}|^2+\mu_0|\partial_\theta \boldsymbol{H}|^2)/\hbar\omega$. This quantity does not reproduce the grating pattern itself, but instead the "effective grating" pattern of the sensitivity fields, again at multiples of the Talbot distance [see the Methods section for details]: with the sources of Fisher information being located at the edges of the real grating apertures, at each self-imaging plane two Fisher information density maxima (rather than one) appear in correspondence to each aperture. It is apparent that grating structures provide a simple means of controlling not only physical field patterns but also sensitivity fields and thus the flow of Fisher information.

Complex diffractive gratings can also generate superoscillatory electromagnetic fields through the interference of multiple coherent waves generated by scattering[34,35]. Superoscillation is a fundamental wave phenomenon whereby, for any fixed spatial bandwidth, finite-energy signals may locally oscillate arbitrarily rapidly. In the context of Maxwell's electrodynamics, this enables the formation of optical hotspots of arbitrarily small size[36] and implies that the resolution of optical microscopy is not fundamentally limited by the conventional Abbe diffraction limit of approximately half the wavelength[37]. Free-space superoscillatory electromagnetic fields are characterised by subwavelength energy localisation, extremely large phase gradients, exceeding the magnitude of the wavevector of the constituent waves, the presence of phase singularities, and regions of energy backflow (retro-propagation). As such, superoscillatory fields closely resemble optical near fields typically encountered in the vicinity of resonant plasmonic nanostructures[38].

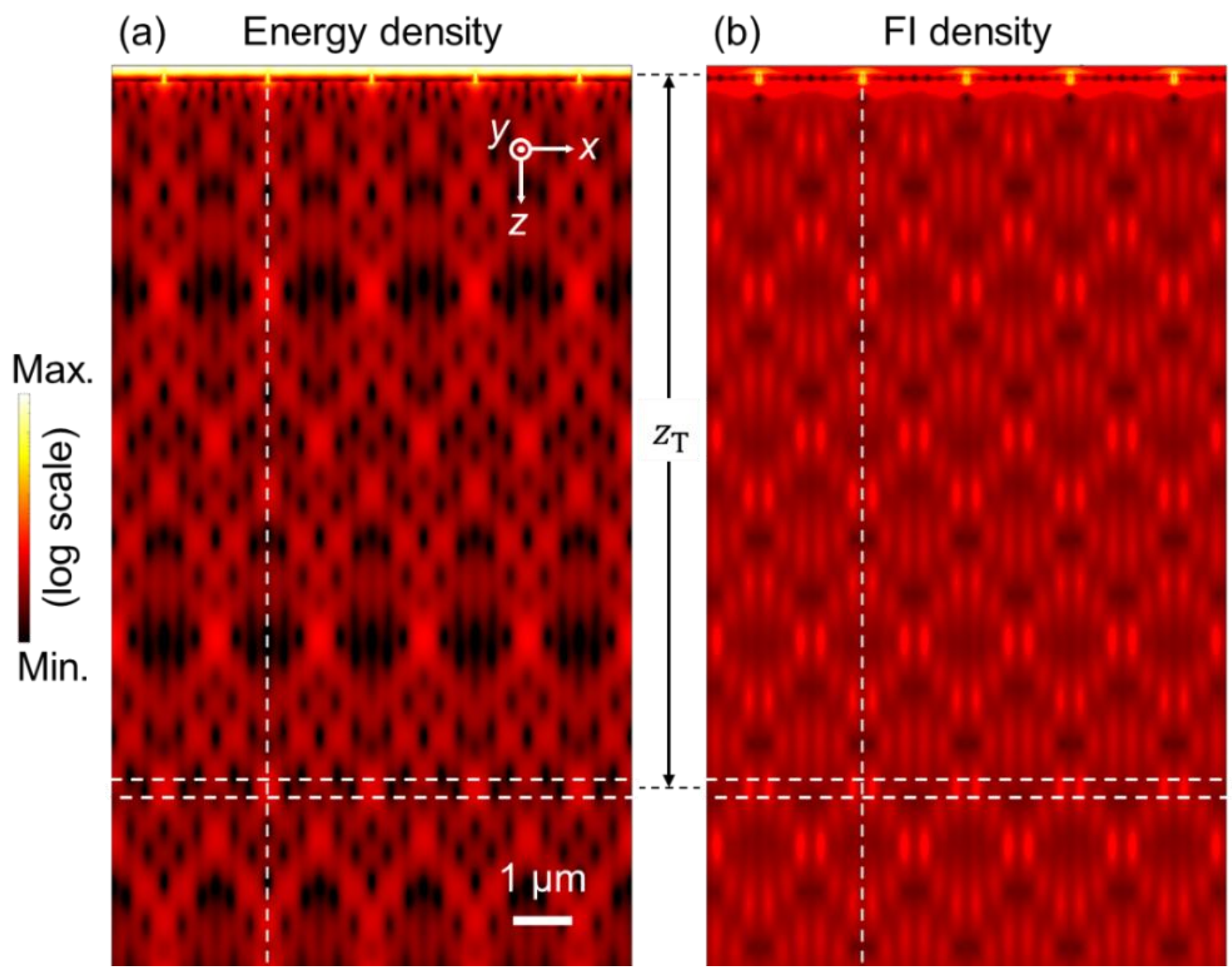

**Fig. 4. Diffraction of information flow: the Fisher information Talbot effect.** (a) The classical optical Talbot effect of periodic 'self-imaging' reconstruction of a grating, shown here for a 150 nm thick gold grating of period $a$ = 1.8 µm with aperture width of 50 nm, under normally-incident, plane wave coherent illumination (from the top) at a wavelength of 505 nm. (b) Fisher information density associated with a 2 nm $x$-displacement of the grating.

Figure 5 compares the concentration of Fisher information in a superoscillatory light field, as would apply to measuring the displacement of a superoscillatory lens (cf. a complex grating) in the transverse $x$-direction, with the case of a conventional parabolic lens. Figures 5a, b, and i show the intensity profiles of fields focused along $\hat{\mathbf{z}}$ by such lenses, assuming plane wave illumination and a numerical aperture of 0.89 in both cases. By design, the superoscillatory field has an intensity distribution in the focal plane of the form[39] $I(x) = 0.18S_2 + S_3$, where $S_i$ are prolate spheroidal wave functions. The full-width half-maxima of the central focal hotspots are $\Delta_c$ = 0.91$\lambda$ for the conventionally focused beam and $\Delta_s$ = 0.33$\lambda$ for the superoscillatory field (i.e. sub-diffractive in the latter case). As reported previously[38], the central energy hotspot of a superoscillatory field is flanked by phase singularities with opposite field vorticity, defined as $\mathbf{\Omega}_P = \nabla \times \mathbf{S}_P$, and zones of energy backflow, where $\mathbf{S}_P$ the Poynting vector (Figs. 5c,d).

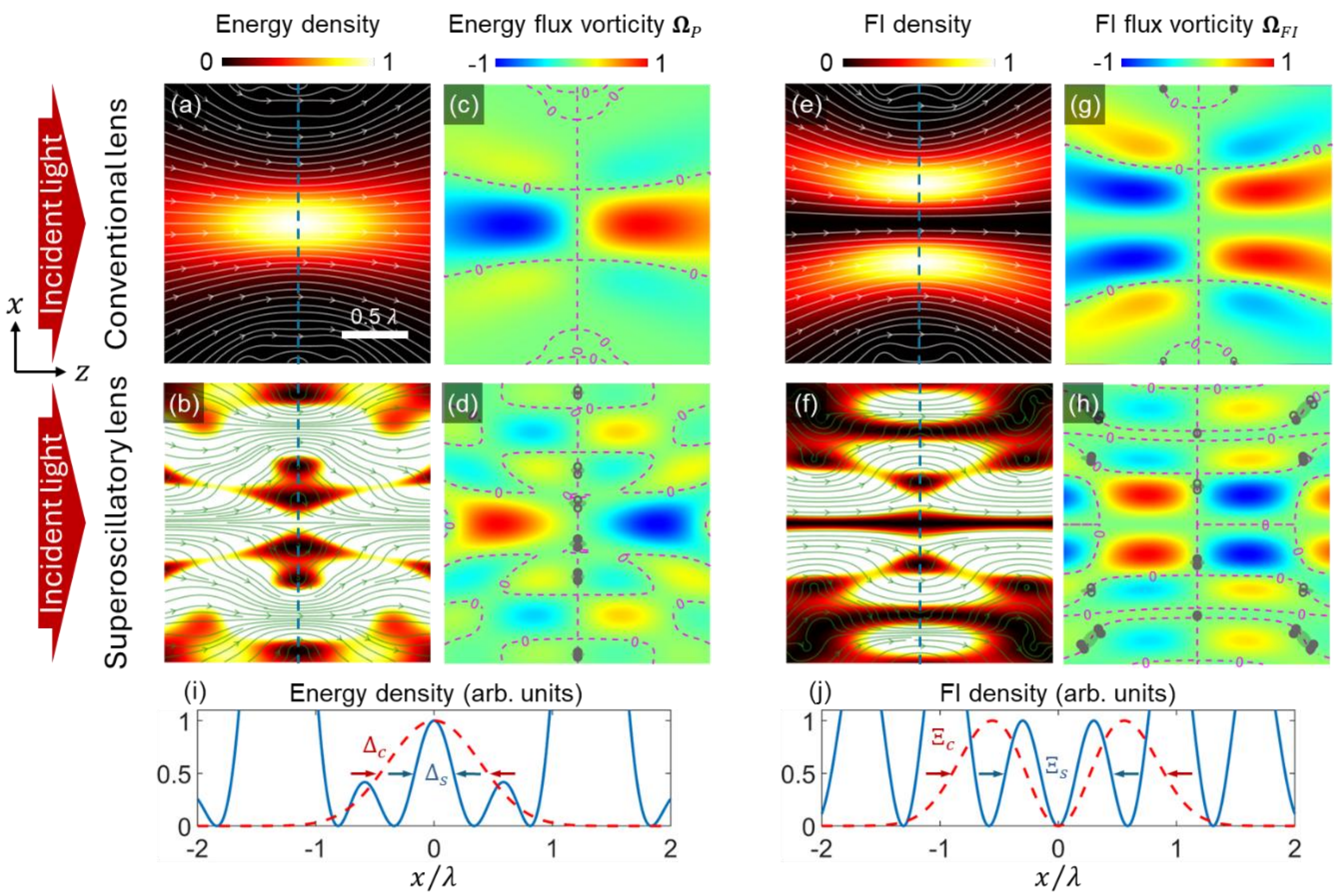

**Fig. 5. Concentration of Fisher information in light focused by conventional and superoscillatory lenses.** (a,b) *x*-*z* cross-sections of the energy density around the foci of (a) conventional and (b) superoscillatory lenses illuminated by a Gaussian beam at a wavelength of 1 μm. Energy density is colour-coded and the direction of the Poynting vector is indicated by overlaid arrows. (c,d) Corresponding maps of energy flow vorticity. (e,f) Fisher information density [color-coded] and flow [arrows] associated with $\Delta x$ = 2 nm displacements of the lenses. (g,h) Corresponding maps of Fisher information flow vorticity. In (c,d,g,h), vorticity centres are indicated by small circles. (i,j) Cross-sectional energy density and Fisher information density profiles along the dashed lines in (a,b) and (e,f).

As for the nanowire position measurements discussed above, destructive interference between Fisher-information sources located on opposite sides of the lenses leads to vanishing Fisher information density and Fisher-information vorticity, $\mathbf{\Omega}_{\mathrm{FI}} = \nabla \times \mathbf{S}_{\mathrm{FI}}$, along the symmetry axes (Figs. 5e–h). Notably, the spatial extent of the central Fisher-information double-peak is smaller for the superoscillatory lens, $\Xi_s = 0.89\lambda$, than for the conventional lens, $\Xi_c = 1.79\,\lambda$ (Fig. 5j).

This example demonstrates that, in close analogy to power focusing, optical Fisher information can be concentrated by a superoscillatory lens to sub-diffractive scales that are unattainable using conventional lenses with the same numerical aperture. This localisation, however, comes at the cost that only a fraction of the total optical power – and correspondingly of the total Fisher information – is channelled through the sub-diffraction region.

While these results demonstrate how to focus Fisher information, they do not address how to design structures that maximise the amount of information generated in the first place. One may ask, what transmission function yields the largest achievable Fisher information about an object's position? In the Methods section, we derive the form of a one-dimensional grating, with transmission $t(x)$ that depends on the position $x$ such that the transmitted wave maximises the Fisher information about $x$ discplacement. For a normally incident plane wave with wavelength $\lambda$, it is found that the optimal periodic grating has a transmission function $t(x) \propto sin\left(\frac{2\,\pi\,x}{\lambda}\right)$, with a period that matches the wavelength $\lambda$. The transmitted waves and sensitivity fields for this structure have a wave vectors, along the plane of the grating. The outgoing plane waves in the optimal configuration thus travel perpendicular to the incident one.  corresponding to a maximal change in momentum along $x$.

### Near field enhancement and vortices of Fisher information

Evanescent sensitivity fields do not transport Fisher information into the far-field. However, Fisher information in evanescent fields can be accessed by decoupling the fields into guided modes, for example, surface plasmon polaritons: Figure 6 considers the lateral displacement (perpendicular to the direction of incident light propagation) of a gold nanocylinder, with plasmonic tapers positioned along the $x$-direction near the nanocylinder, such that surface plasmon polaritons propagate along the taper and are then coupled into the far field by a grating. These excitations are associated with momentum transfer along $x$, carrying significant Fisher information about the $x$-displacement of the nanocylinder. On the other hand, due to the tapers' orientation, the excitations transfer little momentum along $y$ and therefore carry little information about $y$- displacement. Such anisotropic responses may be exploited, e.g., to separate different information channels and to minimize cross talk between them. While providing access to more complete spatial-frequency information, the taper tips also induce local field enhancements that can increase the available Fisher information, by enhancement of the information sources.

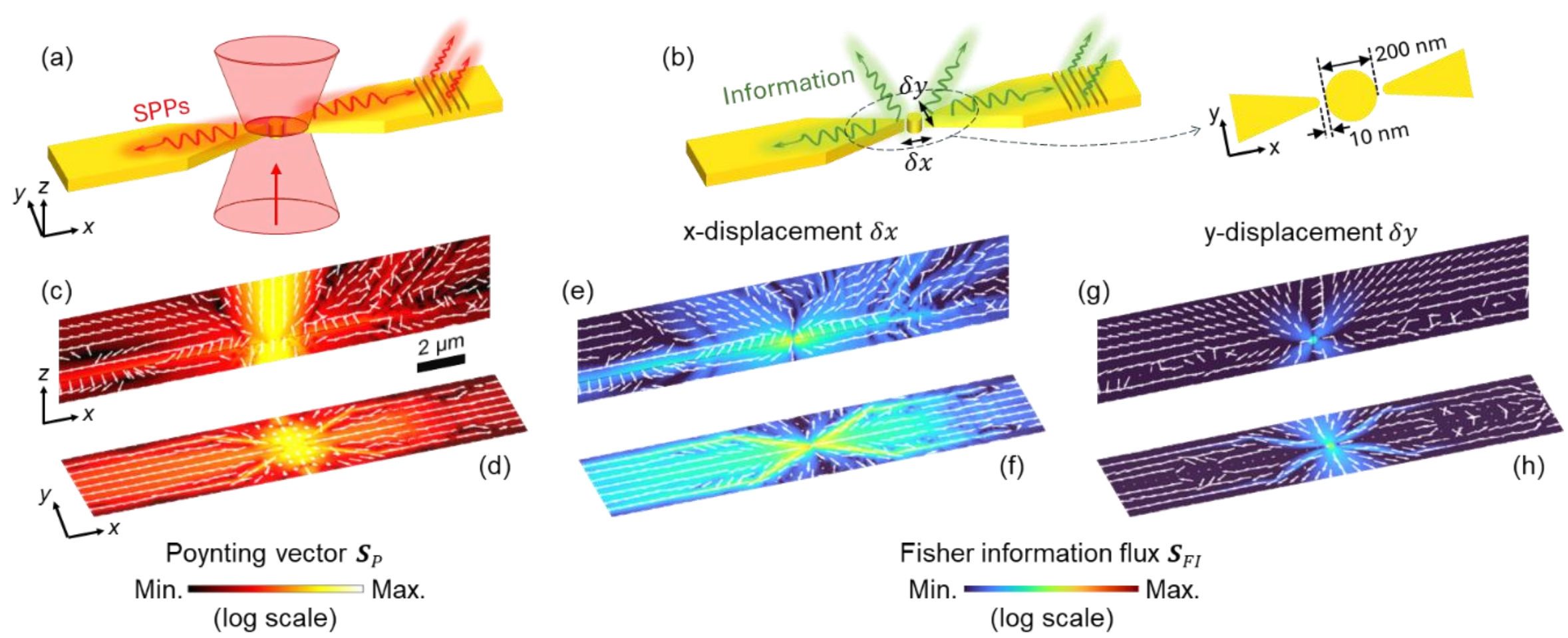

**Fig. 6. Measuring the displacement of a nanoparticle with nearfield probes.** (a,b) A plasmonic nanocylinder [gold, 100 nm diameter, 50 nm thick] located between a pair of plasmonic tapers aligned along $x$ [gap size 10 nm] is illuminated by a focused Gaussian beam incident along $\hat{\boldsymbol{z}}$ [wavelength 644 nm]. A proportion of both the scattered energy and Fisher information associated with displacement of the nanocylinder is coupled to surface plasmon polaritons on the tapers, then scattered into the far field from the taper tips and/or a grating. (c,d) Poynting vector maps in the (c) $x$-$z$ plane at $y = 0$ and (d) $x$-$y$ plane at $z = 5$ nm. (e-h) Corresponding Fisher information flux maps for 2 nm (e,f) $x$- and (g,h) $y$-displacements of the nanocylinder. Color denotes magnitude; overlaid arrows vector directions.

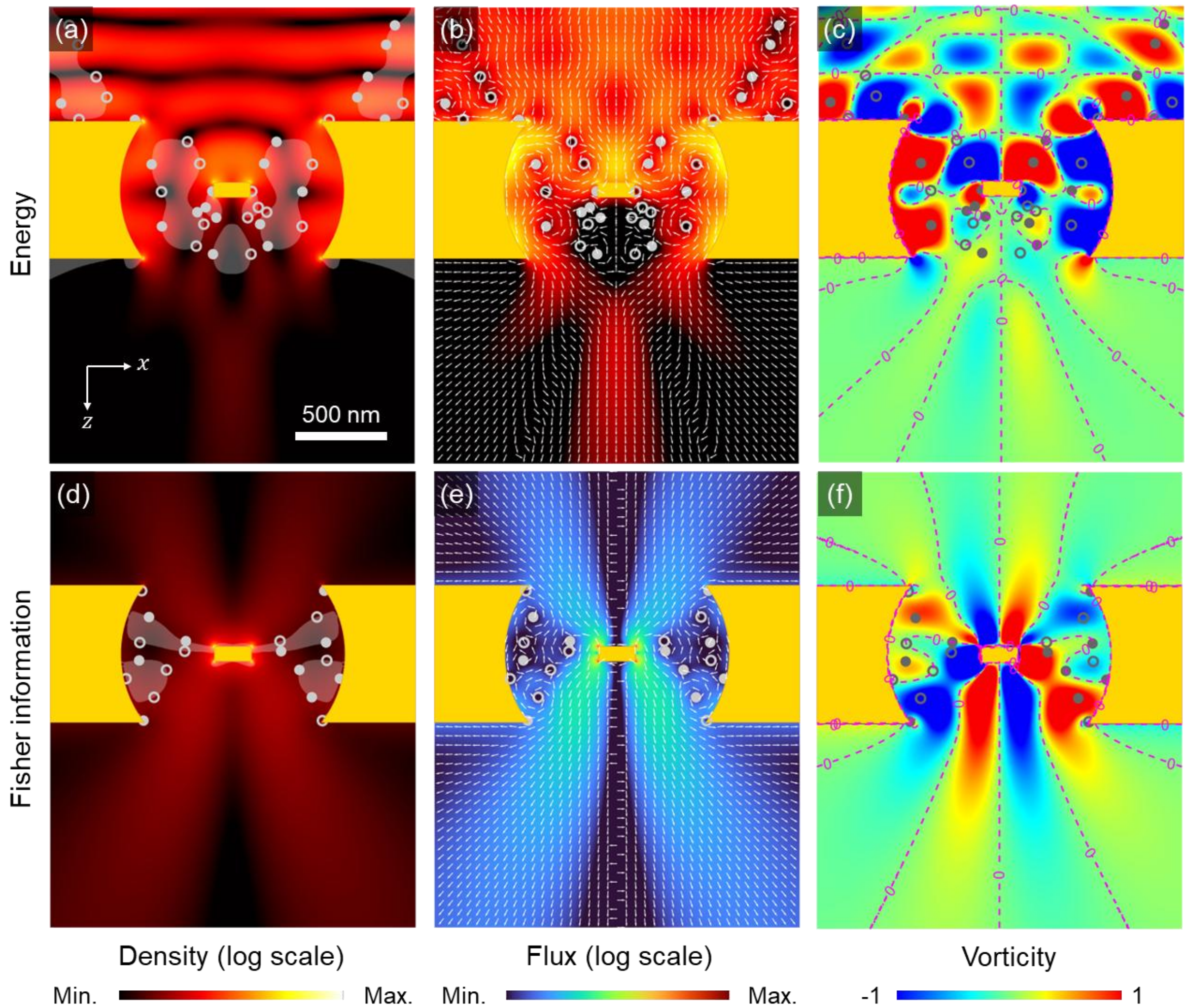

**Fig. 7. Fisher information in a resonant cavity.** Measurements of nanowire displacement [along $x$] in a truncated circular plasmonic cavity under plane-wave illumination along $z$ at a wavelength of 585 nm. (a,b,c) Cross-sectional maps of energy density, flux and flux vorticity. (d,e,f) Corresponding maps of Fisher information density, flux and flux vorticity. Color denotes magnitude; overlaid arrows vector directions. Singular points at centers of clockwise and counterclockwise vorticity are marked with filled and open circles, respectively. Regions of energy and information backflow in (a) and (d) are shaded in grey. Vorticities are defined as $\boldsymbol{\Omega}_P=\nabla\times\boldsymbol{S}_P$ and $\boldsymbol{\Omega}_{FI}=\nabla\times\boldsymbol{S}_{FI}$, with zero-vorticity boundaries in (c) and (f) marked by pink dashed lines.

Placing scattering objects in the near-field of a target is not the only way to enhance the generation and flow of Fisher information. Alternatively, a target may be placed inside a plasmonic cavity with a resonance matched to that of the target, whereby the electromagnetic field is locally enhanced in the vicinity of the target scatterer, increasing the magnitude of the information source and projecting more information into the far field. This is illustrated in Fig. 7 for a plasmonic nanowire in a resonator. Here, we observe an enhancement of total Fisher information by a factor of ~2.5. We also find vortices and regions of backflow in Fisher information flux (Fig. 7b,c), as can occur in optical energy flux[36, 38]. In both cases, the winding directions of vortices are antisymmetric with respect to the $x = 0$ symmetry line. Vortices in the Poynting vector and the Fisher information flow vector do not align spatially, but in two dimensions, singularities (centres of rotation) in both are characteristically pinned to the boundaries of regions where energy and information flow backwards towards their respective sources[38,40,41].

## Conclusions

In this work, we advance the understanding of optical metrology by demonstrating that Fisher information is not merely a statistical measure of estimation precision, but a physically propagating, wave-like quantity. This reinterpretation allows us to treat Fisher information analogously to energy, governed by conservation laws and capable of exhibiting resonance, diffraction, interference, and even backflow and vorticity. Through theoretical analysis and numerical simulation, we have shown that the generation and flow of Fisher information are shaped by the same electromagnetic principles that govern optical fields – yet they differ in key respects, such as directionality, interference behaviour, and spectral response. In general, the Fisher information carried by a light beam does not necessarily propagate in the same way as the optical power.

We have identified the effective sources of Fisher information as arising from spatial regions where optical fields are sensitive to variations in the measured parameter. These give rise to sensitivity fields that can interfere, and under certain conditions, destructively suppress each other, leading to cancellations in the information flow – regardless of where the optical energy propagates. By tailoring the target geometry and its surrounding environment, e.g., through the introduction of resonant plasmonic structures or the careful alignment of wavefronts, we demonstrated the ability to enhance, suppress, or redirect this information flow with high precision.

As key conceptual advances, we introduce how reciprocity manifests for sensitivity fields, which provides a symmetry principle for information exchange in scattering systems and has, like the conventional reciprocity relation, implications on the design of (nano)photonic structures. Moreover, we demonstrate how Fisher information can be engineered and maximised with gratings and resonators and how near-field effects can be harnessed to further control and redirect Fisher information.

Importantly, this framework allows optical metrology, in its various forms[42,43], to be reimagined as an exercise not only in measurement design but in "information field engineering". Instruments can now be constructed to shape the spatial and spectral distribution of information, in the same way as antennas and metasurfaces are used to sculpt electromagnetic energy. This has immediate consequences for the design of ultra-precise sensors, particularly at the nanoscale, where environmental design and interference effects can be leveraged to surpass limitations of traditional imaging.

Beyond the optical metrology of small objects, our findings are applicable to optimising Light Detection and Ranging (LiDAR), remote sensing, and radar technologies. Looking forward, our findings open avenues for the development of "information photonics" – systems where light is not only used to carry energy or data, but where the physical structure of information flow is a design target. By integrating the tools of wave physics, scattering theory, and information geometry, the metrology of the future can become both more precise and more adaptive, with applications in nanotechnology, biomedical imaging, and quantum sensing.

## Acknowledgements

This research was funded in parts by the UK Engineering and Physical Sciences Research Council (EPSRC) [EP/T02643X/1, EP/Z53285X/1] and the Austrian Science Fund (FWF) [10.55776/PIN7240924].

## Author contributions

All authors discussed the theoretical framework, case studies and contributed to writing the manuscript. M.W., L.N. and J.H. performed the analytical calculations with input from S.R. The numerical calculations were carried out by H.Z., L.N., and T.G. with input from K.M., M.W., J.H., S.R and N.Z. The project was proposed and supervised by S.R. and N.Z.

# Controlling the Flow of Information in Optical Metrology: Methods

### Fisher information and scattered power

Here we show how the Fisher information and the scattered power behave for different wavelengths. Our starting point is the simple equation $\partial_\theta^2(\boldsymbol{c}_{scat}^\dagger \boldsymbol{c}_{scat}) = 2(\partial_\theta \boldsymbol{c}_{scat})^\dagger \partial_\theta \boldsymbol{c}_{scat} + 2Re((\partial_\theta^2 \boldsymbol{c}_{scat})^\dagger \boldsymbol{c}_{scat})$, where $\theta$ is the parameter of interest. In this equation we can identify the Fisher information as $J = 4(\hbar\omega)^{-1}(\partial_\theta \boldsymbol{c}_{scat})^\dagger \partial_\theta \boldsymbol{c}_{scat}$ and the scattered power as $P_{scat} = \boldsymbol{c}_{scat}^\dagger \boldsymbol{c}_{scat}$. Rewriting this leads to the following expression: $J = 2(\hbar\omega)^{-1}\partial_\theta^2(P_{scat}) - 4(\hbar\omega)^{-1}Re((\partial_\theta^2 \boldsymbol{c}_{scat})^\dagger \boldsymbol{c}_{scat})$. Importantly, due to the first term only depending on the scattered power this means that the second term contains the full FI contained in the phase and the energy redistribution of the outgoing waves. In the following we will discuss two special cases for the parameter of interest $\theta = x_{pos}$.

<u>Fisher information contained only in the derivative of total scattered power</u>

In the first case, we will investigate a configuration in which the FI is only contained in the first term of the above expression. We assume that a subwavelength particle is located at a node of a standing wave or a radial node of a Bessel beam. As the field is zero at the particle's position, no scattering occurs and the scattered field will be zero, $\boldsymbol{c}_{scat} = \boldsymbol{0}$. In this case the second term in the previous equation becomes zero and the Fisher information is solely given by the second derivative of the scattered power $J = 2(\hbar\omega)^{-1}\partial_\theta^2 P_{scat}$, which is in our case proportional to the spatial curvature of the electric field intensity of the incident field at the location of the particle.

<u>Fisher information contained in total scattered power</u>

For an incoming plane wave travelling along the $z$-axis, the total scattered power is independent of $\theta = x_{pos}$, such that the first term in the above expression becomes zero. We can now analyze the FI using the $T$-matrix, which connects the incident and outgoing wave $\boldsymbol{c}_{scat} = 2T\boldsymbol{c}_{in}$. Using the equation for the Fisher information: $J = -4(\hbar\omega)^{-1}Re((\partial_\theta^2 \boldsymbol{c}_{scat})^\dagger \boldsymbol{c}_{scat})$ , we can rewrite the second derivative as $\partial_x^2 = T(\partial_x^2 D|_{x=0})T^{-1} - 2(\partial_x D|_{x=0})T(\partial_x D|_{x=0})T^{-1} + \partial_x^2 D|_{x=0}$, where $D$ is the translation operator[1]. The translation operator allows to calculate the T-matrix for another position $T(x) = D(-x)T(0)D(x)$. Due to plane wave illumination and displacement orthogonal to this illumination direction the incident field is invariant with respect to these displacements and thus we get $(\partial_x^2 D|_{x=0})T^{-1}\boldsymbol{c}_{scat} = \frac{1}{2}\partial_x^2 D|_{x=0}\boldsymbol{c}_{in} = 0$ and $(\partial_x D|_{x=0})T^{-1}\boldsymbol{c}_{scat} = \frac{1}{2}(\partial_x D|_{x=0})\boldsymbol{c}_{in} = 0$. This implies that the first and second terms vanish in the Fisher information expression and only the second derivative $\partial_x^2 D|_{x=0}$ contributes. In the basis of regular vector spherical wavefunctions, the translation operator can be written using spherical Bessel functions $j_n(k|\boldsymbol{\Delta x}|)$. The derivative only acts on this radial part and after evaluating at $x = 0$ only coefficients with $j_2(k|\boldsymbol{\Delta x}|) \sim (k|\boldsymbol{\Delta x}|)^2$ contribute such that $\partial_x^2 D|_{x=0} \sim k^2$. With that we can write for the Fisher information $J \propto kRe\left(A\boldsymbol{c}_{scat}^\dagger \boldsymbol{c}_{scat}\right)$, where $A$ has the same form as the translation operator, but the spherical Bessel functions are replaced with the second derivative of the spherical Bessel function[2].

In contrast, the scattered power can be written as $P_{scat} = \boldsymbol{c}_{scat}^\dagger \boldsymbol{c}_{scat}$. If the scattering only has a contribution from one mode, we can write $\boldsymbol{c}_{scat} = a_i \boldsymbol{e}_i$, which give us $P_{scat} = |a|^2$ for the scattered power and $J \propto kRe(A_{ii})|a|^2$. With that we can see that Fisher information scales with $J \propto kP_{scat}$ for small scatterers.

**Polarization dependence of Fisher information**

Here we provide a more detailed explanation as to why the Fisher information on different displacement parameters is dependent on the polarization of incident light. Let us start with the simplest case: The scattered field of a dipole scatterer with polarizability $\alpha$ is given by $\boldsymbol{E}_{sc} = \frac{k^2}{4\pi\varepsilon_0}\frac{e^{ik\rho}}{\rho}\hat{\boldsymbol{r}} \times \left(\hat{\boldsymbol{r}} \times \mathbf{p}(\mathbf{r}_0)\right)$ with $\rho = |\boldsymbol{r} - \boldsymbol{r}_0|$ and a dipole moment $\boldsymbol{p} = \varepsilon_0 \alpha E_0 e^{i\boldsymbol{k}\cdot\boldsymbol{r}_0}$ parallel to the polarization of the incident field, which is assumed to be a plane wave with wave vector $\boldsymbol{k}$. The sensitivity field for a parameter of interest $\theta = r_{0_j}$ becomes $\partial_{r_{0_j}}\boldsymbol{E}_{sc} = ik\left(\hat{k}_j - \hat{r}_j\right)\boldsymbol{E}_{sc}(\boldsymbol{r}; \boldsymbol{r}_0 = 0)$ in the far field approximation. This shows that the changes of the scattered field appear only in phase and the sensitivity field is therefore proportional to the scattered field. Under this condition, the Fisher information is $J_{r_{0_j}} \propto \int |\partial_\theta \boldsymbol{E}_{sc}|^2 r^2 d\Omega$, giving $J_{r_{0_j}} \propto \left(5\left(\hat{k}_j\right)^2 - \left(\hat{p}_j\right)^2 + 2\right)$. This

expression shows explicitly how Fisher Information depends on the propagation direction and polarization of incident light. If the incident beam is propagating along the $z$-direction and is polarized along $x$, then the ratios of Fisher information on displacements along $x$:$y$:$z$ are 1:2:7.

This phenomenon can also be considered from the perspective that dipole radiation has a $\sin(\theta)$ dependence and Fisher information radiation is suppressed along the symmetry plane. When the displacement parameter of interest is aligned with the incident polarization, the Fisher information symmetry plane is precisely aligned with the maximum of $\sin(\theta)$. This leads to lower Fisher information than for the other two displacement directions because the symmetry planes for those parameters do not align with the radiated part. Another key point, which increases Fisher information for the case where the displacement direction is aligned with the wave vector of the incoming field, is that the induced dipole changes phase with such a displacement, leading to increased Fisher information (by a factor of 7). This factor is larger than the others, because the whole field experiences a uniform change.

The same arguments can be applied for the magnetic dipole, but with a different polarization. For higher order resonances the same methodology can be used, but the effects are different. For example, the electric and magnetic quadrupoles give the same Fisher information regardless of polarization, due to their $\sin(\theta)\cos(\theta)$ dependence, which means that the symmetry plane for $x$ and $y$ will always be at points where the field is zero.

**Scattering at a nanorod: Fisher information in the forward direction**

Our goal here is to calculate the FI arriving at a small detector in the far-field of a thin rod. This will allow us to find the angles of an incident plane-wave that result in minimal or maximal FI arriving at the detector. From symmetry arguments[3], we already know that a normally incident plane-wave does not carry FI along the forward direction, independent of the material of the rod. To analyze also oblique angles of incidence, we further assume that the scattering is sufficiently weak and employ the Born approximation[4,5], where we approximate the field inside the scattering object by the incident field. This is generally only valid if the refractive index of the object is close to that of its environment. However, we observe that even for metallic objects the angles we predict using these approximations are very similar to those obtained in our numerical simulations. The detector is placed sufficiently far from the object such that only the far-field contributes. Moreover, we assume for simplicity that the polarization of the incident plane-wave and the plane-wave arriving at the detector is the same and we employ a scalar wave approximation. These approximations allow us to directly use results from basic scattering theory. Starting from $\boldsymbol{E}_{scat} \propto \int \mathrm{G}(\boldsymbol{r},\boldsymbol{r}')\Delta\epsilon(\boldsymbol{r}')\boldsymbol{E}_{in}(\boldsymbol{r}')\, d^3\boldsymbol{r}'$, with $\boldsymbol{E}_{in}(\boldsymbol{r}) \propto \mathrm{e}^{i\boldsymbol{k}_{in}\boldsymbol{r}}$ and where $\Delta\epsilon$ is the difference of the permittivity of the object and the environment. We evaluate the field at the point $\boldsymbol{r} = (0,0,d)^{\top}$ with large $d$ and approximate the Green's function at that point by $G(\boldsymbol{r},\boldsymbol{r}') \approx \frac{e^{ik_{in}d}e^{-ik_{in}\boldsymbol{e}_r\boldsymbol{r}'}}{4\pi d}$. We further specify that we want to measure the $x$-position of the object and express the dielectric function as $\epsilon(\boldsymbol{r}) - \epsilon_0 = \Delta\epsilon(\boldsymbol{r}) = \Delta\epsilon_x(x)\delta(y)\delta(z)$, which is justified for a long object aligned with the $x$-direction. For a homogeneous medium we can use $\Delta\epsilon_x(x) \propto \Theta\left(\frac{L}{2} - |x|\right)$, where $L$ is the length of the rod and $\Theta(x)$ is the Heaviside step function. Calculating the scattered field, both for the object $\Delta\epsilon_x(x)$ and a slightly shifted object $\Delta\epsilon_x(x-\delta)$, we find that the sensitivity field for measuring the $x$-position of the rod, evaluated at the point $\boldsymbol{r}$ reads $\partial_\delta \boldsymbol{E}_{scat} \propto \sin\left(\frac{\mathrm{k}_{\mathrm{in,x}}L}{2}\right)$, where $\mathrm{k}_{\mathrm{in,x}} = \mathbf{k}_{\mathrm{in}}\boldsymbol{e}_x = k_{in}\, sin(\alpha)$ and $\alpha$ the angle of incidence, hence $|\partial_\delta \boldsymbol{E}_{scat}(\boldsymbol{r})|^2 \propto sin^2\left(\frac{k_{in} sin(\alpha) L}{2}\right)$. This expression vanishes when $sin\,(\alpha) = \frac{\lambda n}{L}$ and is maximal when $sin\,(\alpha) = \frac{\lambda\left(n+\frac{1}{2}\right)}{L}$.

**Wave equation for the sensitivity fields**

In the following, we show how the wave equation for the sensitivity fields is derived. Starting from Maxwell's equations in the presence of (non-magnetic and time independent) matter, and using the constitutive relations $\boldsymbol{B} = \mu_0\boldsymbol{H}$ and $\boldsymbol{D} = \epsilon\boldsymbol{E}$, it is easy to show that $\nabla \times \nabla \times \boldsymbol{E} + \mu_0\epsilon\partial_t^2\boldsymbol{E} = -\mu_0\partial_t\boldsymbol{j}$ . The derivative with respect to the parameter $\theta$ yields $\nabla \times \nabla \times [\partial_\theta\boldsymbol{E}] + \mu_0\epsilon\partial_t^2[\partial_\theta\boldsymbol{E}] = -\mu_0\partial_t[\partial_\theta\boldsymbol{j}] - \mu_0[\partial_\theta\epsilon]\partial_t^2\boldsymbol{E}$. Introducing the effective electric current $\boldsymbol{j}_{eff}^e = [\partial_\theta\epsilon]\partial_t\boldsymbol{E} + \partial_\theta\boldsymbol{j}$ yields $\nabla \times \nabla \times [\partial_\theta\boldsymbol{E}] + \mu_0\epsilon\partial_t^2[\partial_\theta\boldsymbol{E}] = -\mu_0\partial_t\boldsymbol{j}_{eff}^e$. An equation governing the magnetic sensitivity field $\partial_\theta\boldsymbol{H}$ and more general expressions for magnetic media can be derived analogously, where additional terms depending on the spatial variation of $\epsilon$ and $\mu$ show up. For evaluating terms like $[\partial_\theta\epsilon]\boldsymbol{E}$, we recommend using numerical techniques that involve a smooth rather than an abrupt change of $\epsilon$ at the target's boundary to avoid ill-defined boundary terms. On a more rigorous level, mathematical treatments of such scenarios based on regularization techniques[6] could provide a promising avenue.

### Reciprocity relation for Fisher information

We generalize the Lorentz reciprocity theorem[4] for passive linear systems such that it applies also to the sensitivity fields. For time-harmonic fields and currents the theorem reads $\int \boldsymbol{j}_1 \cdot \boldsymbol{E}_2 - \boldsymbol{j}_2 \cdot \boldsymbol{E}_1 dV = \oint (\boldsymbol{E}_1 \times \boldsymbol{H}_2 - \boldsymbol{E}_2 \times \boldsymbol{H}_1) d\boldsymbol{S}$, where $\boldsymbol{j}_{1/2}$ are the current densities that produce the fields $\boldsymbol{E}_{1/2}$ and $\boldsymbol{H}_{1/2}$. For only outward propagating waves in the far-field, the expression on the right vanishes and we are left with the Lorentz reciprocity theorem in its widely used form $\int \boldsymbol{j}_1 \cdot \boldsymbol{E}_2 - \boldsymbol{j}_2 \cdot \boldsymbol{E}_1 dV = 0$ . To arrive at a similar expression for the sensitivity fields, we first derive Maxwell's equations in the presence of matter with respect to a parameter of interest $\theta$ $\nabla \times \boldsymbol{E} = i\omega \boldsymbol{B} \Rightarrow \nabla \times [\partial_\theta \boldsymbol{E}] = i\omega[\partial_\theta \boldsymbol{B}]$ and $\nabla \times \boldsymbol{H} = -i\omega \boldsymbol{D} + \boldsymbol{j} \Rightarrow \nabla \times [\partial_\theta \boldsymbol{H}] = -i\omega[\partial_\theta \boldsymbol{D}] + \partial_\theta \boldsymbol{j}$. The same is done for the constitutive relations $\boldsymbol{D} = \epsilon \boldsymbol{E} \Rightarrow \partial_\theta \boldsymbol{D} = [\partial_\theta \epsilon]\boldsymbol{E} + \epsilon[\partial_\theta \boldsymbol{E}]$ and $\boldsymbol{B} = \mu \boldsymbol{H} \Rightarrow \partial_\theta \boldsymbol{B} = [\partial_\theta \mu]\boldsymbol{H} + \mu[\partial_\theta \boldsymbol{H}]$. Using the identity $\nabla \cdot (\boldsymbol{A} \times \boldsymbol{B}) = \boldsymbol{B} \cdot (\nabla \times \boldsymbol{A}) - \boldsymbol{A} \cdot (\nabla \times \boldsymbol{B})$ one can verify that $\int [\partial_\theta \boldsymbol{j}_1] \cdot [\partial_\theta \boldsymbol{E}_2] - [\partial_\theta \boldsymbol{j}_2] \cdot [\partial_\theta \boldsymbol{E}_1] dV - \oint [\partial_\theta \boldsymbol{E}_1] \times [\partial_\theta \boldsymbol{H}_2] - [\partial_\theta \boldsymbol{E}_2] \times [\partial_\theta \boldsymbol{H}_1] dS = i\omega \int \partial_\theta \mu ([\partial_\theta \boldsymbol{H}_1] \cdot \boldsymbol{H}_2 - [\partial_\theta \boldsymbol{H}_2] \cdot \boldsymbol{H}_1) dV + i\omega \int [\partial_\theta \epsilon]([\partial_\theta \boldsymbol{E}_2] \cdot \boldsymbol{E}_1 - [\partial_\theta \boldsymbol{E}_1] \cdot \boldsymbol{E}_2) dV$ , where $\omega$ is the oscillation frequency of the time-harmonic currents and fields. In the far-field where only radiating fields contribute, we can again neglect the contributions from the surface integrals and arrive at $\int \boldsymbol{j}^e_{eff,1} \cdot [\partial_\theta \boldsymbol{E}_2] - \boldsymbol{j}^e_{eff,2} \cdot [\partial_\theta \boldsymbol{E}_1] - \boldsymbol{j}^m_{eff,1} \cdot [\partial_\theta \boldsymbol{H}_2] + \boldsymbol{j}^m_{eff,2} \cdot [\partial_\theta \boldsymbol{H}_1] dV = 0$, where we inserted the expressions for the effective electric and magnetic currents $\boldsymbol{j}^e_{eff} = -[\partial_\theta \epsilon] i\omega \boldsymbol{E} + \partial_\theta \boldsymbol{j}$ and $\boldsymbol{j}^m_{eff} = -[\partial_\theta \mu] i\omega \boldsymbol{H}$ .

### Adjoint variable method for Fisher information objectives

In the design of (nano)photonic structures it is common to treat the (monochromatic) scattering problem as a linear system of equations $A\boldsymbol{E} = -i\omega_0 \boldsymbol{j}$, where $A$ is a discretized version of the operator $A = -\nabla \times \frac{1}{\mu} \nabla \times + \omega_0^2 \epsilon$, $\boldsymbol{j}$ is the (discretized) current source and $\boldsymbol{E}$ represents the electric field[7,8]. Similarly, a sensitivity field $\partial_\theta \boldsymbol{E}$ solves the equation $A[\partial_\theta \boldsymbol{E}] + [\partial_\theta A]\boldsymbol{E} = -i\omega_0 \partial_\theta \boldsymbol{j}$.

We now consider a real valued cost function $C = C(\boldsymbol{E}, \partial_\theta \boldsymbol{E}) = \int c(\boldsymbol{E}(\boldsymbol{r}), \boldsymbol{E}^*(\boldsymbol{r}), \partial_\theta \boldsymbol{E}(\boldsymbol{r}), \partial_\theta \boldsymbol{E}^*(\boldsymbol{r}))d\boldsymbol{r}$, which could be, for example, an objective related to the sensitivity field incident on a detector of a specified shape, or a more complicated function involving both, the flow of energy and FI. In the following, $\boldsymbol{\xi}$ denotes design parameters of the scattering system $A$ and possibly also of the current $\boldsymbol{j}$. To perform gradient descent on the cost function, one need to calculate $\partial_\xi C = 2\,Re\,[\left\langle \frac{\partial c}{\partial \boldsymbol{E}}, \partial_\xi \boldsymbol{E} \right\rangle + \left\langle \frac{\partial c}{\partial[\partial_\theta \boldsymbol{E}]}, \partial_\xi [\partial_\theta \boldsymbol{E}] \right\rangle]$, where $\langle u, v\rangle = \int u(r)\cdot v(r)dr$ and we used the identity $\frac{\partial c}{\partial \boldsymbol{E}} = \left(\frac{\partial c}{\partial \boldsymbol{E}^*}\right)^*$ for the real c. Differentiating the linear equations governing $\boldsymbol{E}$ and $\partial_\theta \boldsymbol{E}$ on both sides with respect to $\boldsymbol{\xi}$, gives $\mathcal{A}\partial_\xi \begin{pmatrix} \boldsymbol{E} \\ \partial_\theta \boldsymbol{E} \end{pmatrix} = -(\partial_\xi \mathcal{A}) \begin{pmatrix} \boldsymbol{E} \\ \partial_\theta \boldsymbol{E} \end{pmatrix} - i\omega_0 \partial_\xi \begin{pmatrix} \boldsymbol{j} \\ \partial_\theta \boldsymbol{j} \end{pmatrix}$ for $\mathcal{A} = \begin{pmatrix} A & 0 \\ \partial_\theta A & A \end{pmatrix}$. Introducing the adjoint field and sensitivity field, respectively, $\mathcal{A}^T \begin{pmatrix} \boldsymbol{E}_{adj} \\ \partial_\theta \boldsymbol{E}_{adj} \end{pmatrix} = \begin{pmatrix} \frac{\partial c}{\partial \boldsymbol{E}} \\ \frac{\partial c}{\partial[\partial_\theta \boldsymbol{E}]} \end{pmatrix}$, gives that $\partial_\xi C = -2\,Re\,\langle \begin{pmatrix} \boldsymbol{E}_{adj} \\ \partial_\theta \boldsymbol{E}_{adj} \end{pmatrix}, \begin{pmatrix} \partial_\xi A & 0 \\ \partial_\xi \partial_\theta A & \partial_\xi A \end{pmatrix} \begin{pmatrix} \boldsymbol{E} \\ \partial_\theta \boldsymbol{E} \end{pmatrix} - i\omega_0 \partial_\xi \begin{pmatrix} \boldsymbol{j} \\ \partial_\theta \boldsymbol{j} \end{pmatrix} \rangle$. Note that the transpose is defined with respect to $\langle u, v\rangle$, i.e. $\langle u, Av\rangle = \langle A^T u, v\rangle$. Reciprocity manifests here as $A^T = A$ (i.e. the boundary term vanishes for purely outgoing waves due to reciprocity) and thus $\mathcal{A}^T = \begin{pmatrix} A & \partial_\theta A \\ 0 & A \end{pmatrix}$ showing that the roles for the adjoint fields are switched in the sense that for the normal fields the field influences the perturbed field, while for the adjoint fields it is the other way around. Calculating $\partial_\xi C$ now requires solving Maxwell's equations four times, each time using the same operator $A$: 1) for the physical field $\boldsymbol{E}$ with the physical source $-i\omega_0 \boldsymbol{j}$; 2) for the sensitivity field $\partial_\theta \boldsymbol{E}$, with the effective source $-i\omega_0 \boldsymbol{j}_{eff}$ and $\boldsymbol{j}_{eff} = \partial_\theta \boldsymbol{j} + \frac{\partial_\theta A}{i\omega_0}\boldsymbol{E}$; 3) for the adjoint sensitivity field $(\partial_\theta \boldsymbol{E})_{adj}$, with the effective source $\frac{\partial c}{\partial[\partial_\theta \boldsymbol{E}]}$; and 4) for the adjoint field $\mathbf{E}_{\text{adj}}$, with the effective source $\frac{\partial c}{\partial \boldsymbol{E}} - (\partial_\theta A)(\partial_\theta \boldsymbol{E})_{adj}$ .

**Talbot effect for Fisher information**

Here we consider an effective 2D system with a $a$-periodic grating located at $z = 0$ illuminated by an incident plane wave $\phi(x,z) = Ae^{ikz}$. We characterize the periodic grating using the function $g(x)$, which we split up into Fourier coefficients $c_n$, i.e. $g(x) = \sum_{n\in N} c_n e^{i\frac{2\pi n}{a}x}$. At $z = 0$ the field $\mathbf{E}(\mathrm{x},\mathrm{z}) = \mathrm{U}(\mathrm{x},\mathrm{z})\mathbf{e}_\mathrm{y}$ is given by $U(x, z = 0) = \phi(x, z = 0)g(x)$ and using the Fresnel approximation for the propagating field we get $U(x,z) = A\frac{e^{ikz}}{\sqrt{iz\lambda}} \sum_{n\in N} \int_{-\infty}^{\infty} c_n e^{i\frac{2\pi n}{a}x'} e^{\frac{ik}{2z}(x-x')^2} dx'$. Solving this integral leads to the expression $U(x,z) = Ae^{ikz}\sum_{n\in N} c_n e^{i\frac{2\pi n x}{a}} e^{-i(\frac{2\pi n}{a})^2\frac{z}{2k}}$, which gives us the diffracted field. To obtain the Fisher information flow we need to calculate the sensitivity field $\partial_\theta U(x,z)$. Depending on the parameter of interest the derivative can act on different parts of this equation, but if we assume the parameter of interest is neither wavelength nor the periodic length of the grating, we can see that the derivative only acts on the Fourier coefficients $c_n$, therefore creating a diffraction pattern for an effective grating. Assuming the parameter of interest to be $\theta = x_{pos}$, the sensitivity field can be calculated using an effective grating given by $\partial_x\big(g(x)\big) = \sum_{n\in N} c_n' e^{i\frac{2\pi n}{a}x}$ with $c_n' = \frac{i2\pi n}{a} c_n$.

**Engineering objects for maximizing Fisher information**

We consider an incident field $\mathbf{E}_{\mathrm{in}} = \mathbf{E}_0 \mathrm{e}^{\mathrm{i}kz}\mathbf{e}_\mathrm{y}$ impinging on a grating $t(x)$ at $z = 0$. The field after the grating is given by $\mathbf{V}(\mathrm{x}) = \mathrm{t}(\mathrm{x})\mathrm{E}_0\mathbf{e}_\mathrm{y}$. Using the angular spectral method[9] for $z > 0$ gives $\mathbf{E}(\mathrm{x},\mathrm{z}) = \int \mathrm{V}(\mathrm{k_x})\mathrm{e}^{\mathrm{i}(\mathrm{k_x x}+\mathrm{k_z z})}\mathrm{dk_x}\,\mathbf{e}_\mathrm{y} = \mathrm{U}(\mathrm{x},\mathrm{z})\mathbf{e}_\mathrm{y}$, where $\mathrm{k_z} = \sqrt{(\mathrm{k}^2 - \mathrm{k_x^2})}$ for $\mathrm{k}^2 \geq \mathrm{k_x^2}$ and $\mathrm{k_z} = \mathrm{i}\sqrt{(\mathrm{k_x^2} - \mathrm{k}^2)}$ otherwise. The Poynting vector is then given by $\mathbf{S}(\mathrm{x},\mathrm{z}) = \mathrm{Re}(\mathbf{E}^* \times \mathbf{H})/2 = (1/(2\mu_0\omega))\mathrm{Im}\Big(\mathrm{U}^*\mathbf{e}_\mathrm{y} \times \big(\nabla \times (\mathrm{U}\mathbf{e}_\mathrm{y})\big)\Big) = (1/(2\mu_0\omega))\mathrm{Im}(U^*\nabla\mathrm{U})$. Integrating the Poynting vector over a plane in the far field: $\mathrm{I} = \int_{-\infty}^{\infty} \mathbf{S}(\mathrm{x},\mathrm{z}) \cdot \mathbf{e}_\mathrm{z}\mathrm{dx} = 1/(2\mu_0\omega)\int_{-\infty}^{\infty} \mathrm{Im}(U^*\,\partial_\mathrm{z}\mathrm{U})\,\mathrm{dx} = 1/(4\mu_0\omega)\int_{-k}^{k} V^*(k_x)V(k_x')\int_{-\infty}^{\infty} e^{i\left((k_x-k_x')x+(k_z-k_z')z\right)}(k_z + k_z')\,\mathrm{dx} =$

$\pi/(\mu_0\omega)\int_{-k}^{k} |V(k_x')|$^2 $\mathrm{k_z dx}$, where it is assumed that the distance between the object and the plane is large, such that evanescent terms $\mathrm{k_x} > \mathrm{k}$ can be neglected. Thus, $\mathrm{I} \approx \int_{-\mathrm{k}}^{\mathrm{k}} \mathrm{k_z}|\mathrm{V}(\mathrm{k_x})|^2\mathrm{dk_x}$. For displacements along the $x$-direction, $\partial_\theta \mathrm{V} = -\partial_\mathrm{x}\mathrm{V}$, radiated Fisher information is given by $\mathrm{FI} \approx \int_{-k}^{k} k_z|\mathcal{F}[\partial_\theta V](k_x)|^2 dk_x = \int_{-\mathrm{k}}^{\mathrm{k}} \mathrm{k_z k_x^2}|\mathrm{V}(\mathrm{k_x})|^2\mathrm{dk_x}$. Optimizing the ratio between FI and the

radiated intensity yields $\mathrm{FI/I} \propto \int_{-k}^{k} k_z k_x^2 |V(k_x)|^2 dk_x \,/\, \int_{-k}^{k} k_z |V(k_x)|^2 dk_x \leq k^2$, where $k^2$ can be approached arbitrarily closely by concentrating $V(k_x)$ at $k = |k_x|$. In physical experiments, one would choose $|k_x| < k$ in such a way that the field can propagate in $z$ direction with acceptable longitudinal component $k_z$.

**Numerical modelling**

In all cases, the refractive index of silicon it taken to be 3.8 (no imaginary part), and optical parameters for gold are taken from Johnson & Christy[10].

The spherical particle numerical simulations presented in Fig. 2 are performed using the optical tweezer toolbox (OTT)[11].

Figure 5 is generated by simulating the propagation of axisymmetric optical fields in free space, using the Hankel-transform-based angular spectrum method[12] (in MATLAB). Fields are defined in the focal plane – the superoscillatory field being described by band-limited circular prolate spheroidal wave functions (CPSWFs), $I(x) = 0.18S_2 + S_3$ – and numerically back- and forward-propagated to obtain complete intensity and phase distributions. Energy and Fisher information flow maps are then derived from these field distributions.

All other numerical simulations are performed using Lumerical FDTD Solutions. Incident light is (unless otherwise stated) an $x$-polarized monochromatic plane wave. Perfectly matched layer (PML) boundary conditions are employed in the case of nanowire displacement simulations (Figs. 3 and 7), while periodic boundary conditions are applied in relation to the Talbot effect (Fig. 4).

Figure 6 assumes a three-dimensional geometry in which a plasmonic gold nanocylinder, 100 nm in diameter and 50 nm thick, is positioned between two plasmonic tapers aligned along the x direction. Each taper has a thickness of 50 nm and consists of a tapered tip and a semi-infinite 3 μm wide waveguide section. The taper tip has an opening angle of ~53° and an apex with a 5 nm radius of curvature. A grating is incorporated on one waveguide, with a period of 618 nm (corresponding to the surface plasmon polariton wavelength), a depth of 25 nm, and duty cycle of 0.4. The structure is illuminated by a Gaussian beam at a wavelength of 644 nm, propagating in the -$z$ direction, focused to a 2 μm half-maximum diameter centered on the nanocylinder.

Figure 7 assumes, and shows in $x$-$z$ cross section, an infinitely long (in the $y$ direction) gold nanowire, 200 nm wide and 80 nm thick, located in a plasmonic cavity. The cavity geometry is formed by removing a cylindrical volume of $x$-$y$ plane radius 615 nm (and infinite $y$ direction length) from a 750 nm thick ($z$) gold film. Energy backflow is defined by the condition $\mathbf{n}_P \cdot \mathbf{S}_P < 0$, where $\mathbf{n}_P$ denotes the direction from the source to a given point, and $\mathbf{S}_P$ is the energy flux. The same criterion naturally extends to the information flux, yielding $\mathbf{n}_{\mathrm{FI}} \cdot \mathbf{S}_{\mathrm{FI}} < 0$. In the configuration presented in Fig. 7, lateral displacement of the nanowire creates a pair of Fisher-information sources at its left and right ends, with equal strength and opposite phase. Owing to the symmetry of the surrounding structure, the information radiation direction $\mathbf{n}_{\mathrm{FI}}$ can be approximated by a vector pointing from the centre of the nanowire to the observation point, allowing information backflow regions to be identified.